\documentclass[twocolumn,prl,aps,showpacs]{revtex4-1}
\usepackage{graphicx}
\usepackage{color}
\usepackage{dcolumn}
\usepackage{amsmath}

\newlength{\figwidth} 
 \newlength{\figwidthb} %
\newcommand{\SIO}{Sr$_3$Ir$_2$O$_7$ }
\newcommand{\SIOns}{Sr$_3$Ir$_2$O$_7$} 
\newcommand{\SIOs}{Sr$_2$IrO$_4$ } 
\newcommand{\SIOsns}{Sr$_2$IrO$_4$} 
\newcommand{\jeff}{$J_{\textrm{eff}}$}

\begin{document}

\title{Giant Magnon Gap in Bilayer Iridate \SIO: Enhanced Pseudo-dipolar
  Interactions Near the Mott Transition} 

\author{Jungho Kim,$^1$ A. H. Said,$^1$ D. Casa,$^1$ M. H. Upton,$^1$
  T. Gog,$^1$ M. Daghofer,$^2$\\ G. Jackeli,$^3$ J. van den Brink,$^2$
  G. Khaliullin,$^3$ B. J. Kim$^4$} 
\address{$^1$Advanced Photon Source, Argonne National Laboratory, Argonne,
  Illinois 60439, USA}
\address{$^2$Institute for Theoretical Solid Sate Physics, IFW Dresden,
  Helmholtzstr. 20, 01069 Dresden, Germany} 
\address{$^3$Max Planck Institute for Solid State Research, Heisenbergstra\ss
  e 1, D-70569 Stuttgart, Germany} 
\address{$^4$Materials Science Division, Argonne National Laboratory, Argonne,
  IL 60439, USA} 

\date{\today}

\begin{abstract}
Using resonant inelastic x-ray scattering, we observe in the bilayer iridate \SIOns, a spin-orbit coupling driven magnetic insulator with a small charge gap,  a magnon gap of $\approx$92 meV for both acoustic and optical branches. This exceptionally large magnon gap exceeds the total magnon bandwidth of $\approx$70 meV and implies a marked departure from the Heisenberg model, in stark contrast to the case of the single-layer iridate \SIOsns. Analyzing the origin of these observations, we find that the giant magnon gap results from bond-directional pseudo-dipolar interactions that are strongly enhanced near the metal-insulator transition boundary. This suggests that novel magnetism, such as that inspired by the Kitaev model built on the pseudo-dipolar interactions, may emerge in small charge-gap iridates.
\end{abstract}


\maketitle

Identifying the hierarchy of energy scales associated with multiple interacting degrees of freedom is the starting point for understanding the physical properties of transition-metal oxides (TMOs). In most TMOs, the largest energy scale is the Coulomb interaction $U$, which suppresses charge motion in Mott insulators and allows description of the low-energy physics in terms of the remaining spin and orbital degrees of freedom.  
In 5$d$ iridium oxides, however, $U$ is significantly diminished due to the spatially extended 5$d$ orbitals, and the correlated insulating state cannot be sustained without the aid of large spin-orbit coupling ($\sim$0.5 eV)~\cite{bjkim08}. This additional interaction competes with other energy scales such as the crystal field and the hopping amplitude. The resulting charge gap is much smaller than that in a typical 3$d$ TMO or even those in most semiconductors, being on the order of 0.1 eV or even smaller~\cite{moon09,moonPRL09}. On the other hand, the energy scale of the magnetic interaction has been recently found to be of the same order of magnitude as that of 3$d$ TMOs~\cite{KimPRL12,FujiyamaPreprint12}. As a consequence, an intriguing new hierarchy may result in which the energy scales for magnetic degrees of freedom surpass that for charge degrees of freedom, ushering in a new paradigm for the magnetism in 5$d$ TMOs.

Iridates of the Ruddlesden-Popper series Sr$_{n+1}$Ir$_n$O$_{3n+1}$ display a systematic electronic evolution as a function of the number of IrO$_2$ layers ($n$); as $n$ increases, the electronic structure progresses toward a metallic ground state as evidenced by the softening of the charge gap in \SIO ($n$=2) and the metallic ground state found for SrIrO$_3$ (n=$\infty$)~\cite{moonPRL09}. The charge gap becomes so small already at $n$=2 that it cannot be resolved even in the optical conductivity spectrum, indicating proximity to the Mott transition point. Thus, the bilayer compound \SIO provides a platform for investigating the nature of magnetism in the small $U$ region near the metal-insulator transition (MIT) boundary. 

In this Letter, we report the magnetic excitation spectra of \SIO measured by
resonant inelastic x-ray scattering (RIXS)~\cite{ament11note}, which show a number of
features characterizing the unconventional nature of the magnetism in \SIO
lying close to a Mott critical point. We observe two anomalous features: a
giant magnon gap of $\approx$92 meV, even larger than the total magnon bandwidth
$\approx$70 meV, which demonstrates a marked departure from the Heisenberg
model; and a very small bilayer splitting ($\approx$5 meV), which is
surprising in view of the ``cubic" shape of the spin-orbit entangled wavefunction in iridates~\cite{bjkim09, jackeli09}, which would suggest strong inter-layer interactions. The observed small bilayer splitting indicates frustration of the inter-layer isotropic exchange interactions. 
The temperature scale of the magnon gap exceeds 1000 K, indicating that the melting of the $G$-type collinear antiferromagnetic (AF) order at $\approx$285 K~\cite{327RXSPaper} is not driven by thermal fluctuations of magnetic moments, but rather by thermal charge carriers. 
Our analysis shows that the large magnon gap results from enhanced pseudo-dipolar (PD) interactions, which has an intriguing implication for the Kitaev model~\cite{Kitaev} discussed recently in the context of honeycomb lattice iridates $A$$_2$IrO$_3$ ($A$=Li or Na) in which the PD interactions play the major role~\cite{jackeli09,chaloupka10,SinghPRB10,HillPRB11,SinghPRL12,ChoiPRL12,YePRB12}.  

\begin{figure*}[t]
\hspace*{-0.1cm}\vspace*{-0.2cm}\centerline{\includegraphics[width=1.9\columnwidth,angle=0]{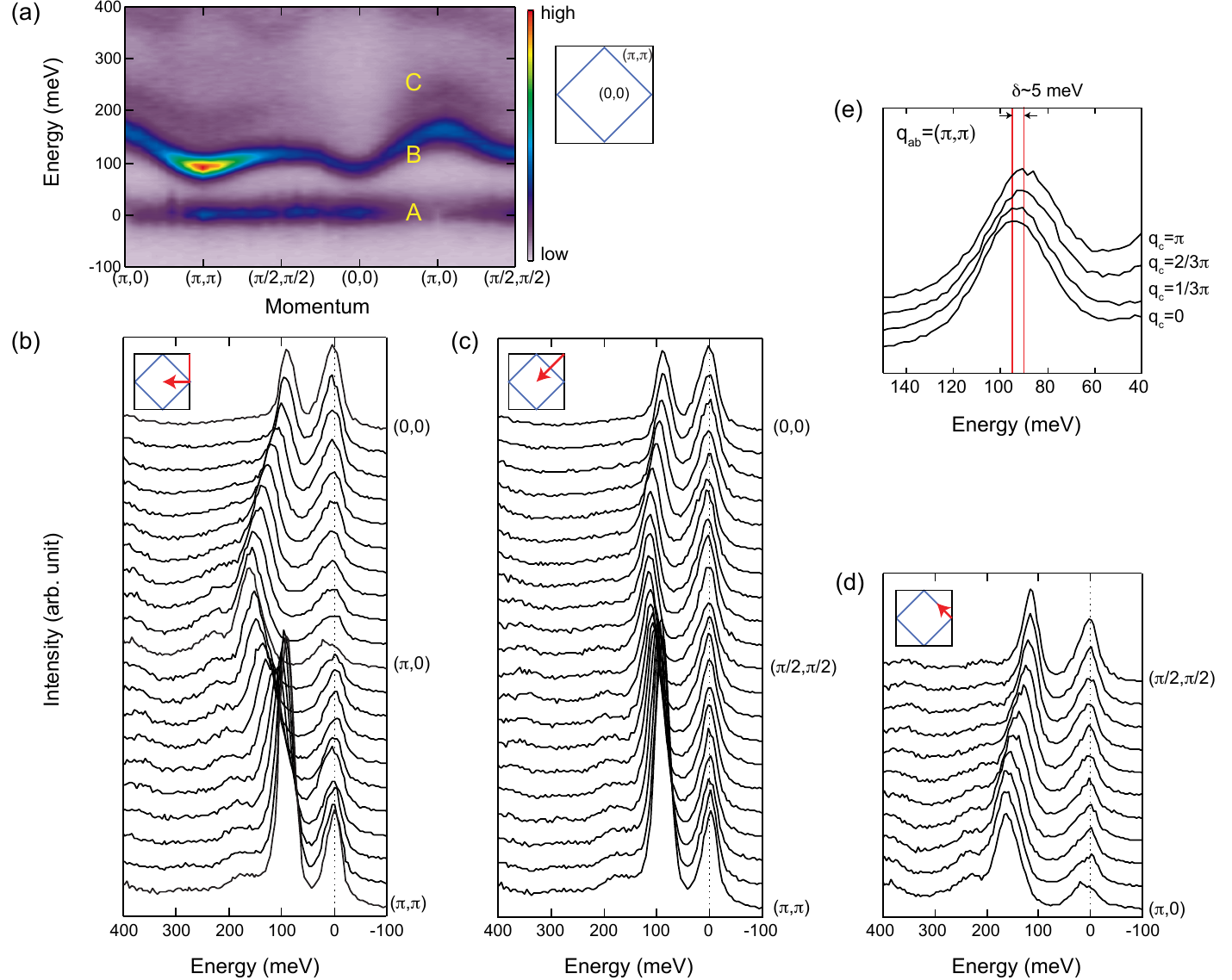}}
\caption{(a) Image and (b-d) stack plot of the RIXS data recorded with $q_{ab}$ along high symmetry lines and $q_c$ fixed at $\frac{\pi}{3}$. Brillouin zone of the undistorted tetragonal unit cell (black square) and the magnetic cell (blue square) is shown with the notation following the convention for the tetragonal unit cell, as, for instance, in La$_2$CuO$_4$. (e) RIXS spectra measured at four different $q_c$'s with $q_{ab}$ fixed at ($\pi$,$\pi$). The $q_c$'s of 0, $\pi$/3, 2$\pi$/3, and $\pi$ correspond approximately to $l$=25.65, 26.5, 27.35, and 28.25, respectively. The red lines indicate the approximate peak positions at $q_c$=0 and $\pi$. }\label{fig:fig1}
\end{figure*}

Experiments were carried out at the 30-ID beamline at the Advanced Photon Source. A horizontal scattering geometry was used with $\pi$-polarized incident photons tuned to Ir L$_{3}$ edge. A spherical diced Si(844) analyzer was used. The overall energy and momentum resolution of the RIXS spectrometer was about 30 meV and $\pm$0.032 $\AA^{-1}$, respectively.  By using a high resolution monochromator and improving the quality of the analyzer, the energy resolution has been improved by more than a factor of four since we recently reported RIXS measurement on the single layer iridate \SIOsns~\cite{KimPRL12}.

Figure 1(a) and 1(b)-(d) show the image and stack plots, respectively, of the
RIXS spectra recorded with in-plane momentum transfer $q_{ab}$ along high
symmetry lines and $q_c$ fixed at $\frac{\pi}{3}$. Three main features in the
spectra are: (A) elastic/quasi-elastic peaks near the zero energy, (B) an
intense and dispersive band in the range $90-160$ meV, and (C) a rather weak
and broad feature above the dipersive band suggestive of two-magnon states. By
comparing to the recent RIXS data~\cite{KimPRL12} and theoretical study~\cite{amentprb11} on the single layer
\SIOsns, it is evident that feature B with similar overall
energy scale and momentum dependence of the intensity (see also Fig.~2) is a
single magnon excitation. With this assignment, however, two anomalies are
apparent: the acoustic branch 
appears to be absent, and the magnon gap is unusually large even for an optical
mode. Typically, two branches of magnetic modes, acoustic and optical, are
observed in other bilayer systems such as bilayer
manganites~\cite{PerringPRL01} and cuprates~\cite{ReznikPRB96}. To identify
the acoustic branch, we scanned along the $q_c$ direction fixing
$q_{ab}$=($\pi$,$\pi$) where the maximal bilayer splitting is expected, as
shown in Fig.~1(e). At $q_c$=$\pi$ (0), only the acoustic (optical) branch has
finite intensity, in accordance with the intuitive notion that magnons emanate
from magnetic Bragg spots. At intermediate $q_c$, the spectrum is contributed
to by both branches with a gradual shift in spectral weight from one branch to
the other.  We see an upward shift of $\approx$5 meV as $q_c$ is varied from $\pi$ to
0. Thus, it is seen that the acoustic branch also has a large gap and is
nearly degenerate with the optical branch. The small splitting of the two
branches implies strongly frustrated inter-layer interactions, which, at first sight,
seems inconsistent with the observed spin-flop transition driven by the
inter-layer interactions~\cite{327RXSPaper}. 

To unravel this paradox, we first determine the origin of the anomalously large magnon gap. Such a large gap signals a marked departure from the Heisenberg model and that the magnetism in bilayer \SIO is therefore qualitatively different from its single layer variant \SIOsns. A recent resonant x-ray diffraction study~\cite{327RXSPaper} establishes that \SIO has a $c$-axis collinear structure, unlike the single layer \SIOs with in-plane canted moments~\cite{bjkim09}. These different magnetic anisotropies in \SIOs and \SIO were captured in a magnetic exchange Hamiltonian derived from microscopic interactions, which we will use here as well, adding to it longer-range interaction terms, which
were also needed in the single layer \SIOs to quantitatively account for the magnon dispersion~\cite{KimPRL12}. The resulting model contains intra-layer and inter-layer interactions; the intra-layer interactions read

\begin{align}
H_{ab} &= \sum_{\langle i,j\rangle} \Bigl[
J\vec{S_i}\vec{S_j}
+\Gamma S^z_iS^z_j
+  D\bigl(S_i^xS^y_j-S_i^yS^x_j\bigr)\Bigr]\nonumber\\
&\quad\quad+ \sum_{\langle\langle i,j\rangle\rangle} J_2\vec{S_i}\vec{S_j}
+ \sum_{\langle\langle\langle i,j\rangle\rangle\rangle} J_3\vec{S_i}\vec{S_j}\;,
\end{align}

\noindent
where $\langle i,j\rangle$, $\langle\langle i,j\rangle\rangle$, and
$\langle\langle\langle i,j\rangle\rangle\rangle$ denote first, second and
third nearest neighbors within each plane, and $J$, $J_2$ and $J_3$ represent
the corresponding isotropic coupling constants [see Fig. 3(a)]. The 
anisotropic coupling $\Gamma$ includes PD terms driven by Hund's exchange 
and those due to staggered rotations of octahedra~\cite{jackeli09}.
The latter also induce a Dzyaloshinsky-Moriya (DM) interaction, with the
corresponding coupling constant $D$. Analogously, the inter-layer 
interactions read

\begin{align}
H_{c} &= \sum_{i} \Bigl[
J_c\vec{S_i}\vec{S}_{i+z}
+\Gamma_c S^z_iS^z_{i+z} \nonumber \\
&\quad +  D_c\bigl(S_i^xS^y_{i+z}-S_i^yS^x_{i+z}\bigr)\Bigr]
+\sum_{\langle i,j\rangle}J_{2c}\vec{S_i}\vec{S}_{j+z}\;, 
\end{align}

\noindent
where the first sum runs over all sites in one plane, and the second
over all next-nearest-neighbor pairs across the planes [see Fig.~3(a)]. 
Inter-layer interactions $J_c$, $\Gamma_c$, and
$D_c$ for nearest-neighbors along $c$ are complemented by an
inter-layer next-nearest--neighbor coupling $J_{2c}$.

In this model, the magnon dispersions are given by

\noindent
\begin{align}\label{eq:omega}
\omega_{\pm}({\bf q}) = S\sqrt{A^2_{\pm}({\bf
q})-X^2_{\pm}({\bf q})-Y^2_{\pm}({\bf q})}
\end{align}
with
\begin{align}
A_{\pm}({\bf q}) &= 4(J+\Gamma) +(J_c+\Gamma_c)- 4 J_2(1-\cos q_x\cos q_y) \nonumber\\
&\quad - 4J_3(1-\gamma_{2{\bf q}})  -
4J_{2c}(1\mp\gamma_{\bf q})\;, \label{eq:Aq}\\
X_{\pm}({\bf q})&= 4J\gamma_{\bf q} \pm J_c,\
Y_{\pm}({\bf q})= 4D\gamma_{\bf q} \pm D_c\;, \label{eq:Yq}
\end{align}

\noindent
where the upper (lower) sign refers to
optical (acoustic) branches, and 
$\gamma_{\bf q} = \frac{1}{2}(\cos q_x + \cos q_y)$. 

%
%
\begin{figure}[t]
\hspace*{-0.2cm}\vspace*{-0.1cm}\centerline{\includegraphics[width=0.8\columnwidth,angle=0]{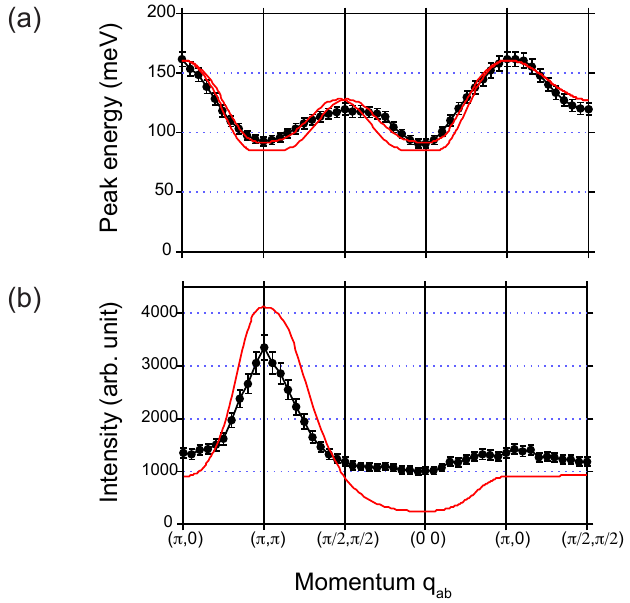}}
%
\caption{Momentum dependence of the (a) position and (b) intensity of the magnon peak extracted by fitting the RIXS spectra (dots with error bar), overlaid with the fit from theory model (red solid lines). Because the acoustic and optical branches are not resolved in the experiment, the intensities for the two branches are summed in the theoretical calculations with appropriate weights (See {\it Supplementary Material} for details) to compare to the experiment.}\label{fig:fig2}
\end{figure}

Following Ref.~\cite{jackeli09}, one can express all of the coupling constants for the isotropic and anisotropic exhange interactions, except the ones for the longer-range interactions ($J_2$,$J_3$ and $J_{2c}$), 
in terms of the three microscopic parameters $\eta$, $\theta$, and $\alpha$, and thus fit the magnon spectrum using these parameters. Here, $\eta$=$J_H$/$U$ is the ratio between Hund's coupling and Coulomb correlation, $\theta$ parametrizes the degree of tetragonal distortion, and $\alpha$ is the octahedra rotation angle. Note that the PD interactions in the strong SOC limit are scaled by $\eta$~\cite{jackeli09}, and $\theta$ describing the deviation from the cubic wavefunction (\jeff=1/2) and $\alpha$ are directly relevant to hopping amplitudes and therefore the superexchange interactions. The expressions for the dependences of the coupling constants on these parameters are provided in the {\it Supplementary Material}. This approach greatly reduces the number of adjustable parameters. Furthermore, experimentally $\alpha$ is determined to be $\approx$$12^\circ$~\cite{Subramanian94,cao02}, and $\theta$ is constrained to be in the range between 0.12$\pi$ and 0.26$\pi$~\cite{327RXSPaper}. 

Figure 2 shows the 
fit of the experimental dispersion and intensity using the above model. We
find that the large gap can only be reproduced when both $\eta$ and $\theta$
are large, about 0.24 and 0.26$\pi$, respectively. Physically, the large
$\eta$ can be understood as arising from the screening of $U$ as the system
approaches the borderline of an MIT, which is evident from the optical data
showing softening of the charge gap~\cite{moon09}. It is well known that while
$U$ is screened in the solid, $J_H$ is not~\cite{AntonidesPRB79,BoschSSC80},
so that the metallic screening results in enhanced $\eta$. The need for the
large $\theta$ can be seen from the fact that when $\theta$ is larger than
$\theta_c$($\approx0.25\pi$), both in-plane ($\Gamma$) and out-of-plane
($\Gamma_c$) PD terms favor the $c$-axis moment~\cite{327RXSPaper} and thus
there is a strong preference for the $c$-axis moment. This pronounced magnetic
anisotropy is amplified by the large $\eta$, which leads to the sizable
gap. While DM terms also contribute to the stabilization of the $c$-easy axis
structure~\cite{327RXSPaper}, their effects in the magnon dispersion are much
smaller than those from the PD terms. With this we find the resulting magnetic exchange interactions shown in Table 1.  In addition to very large anisotropic couplings, we find that the nearest-neighbor Heisenberg coupling $J$ is also enhanced compared to its measured~\cite{KimPRL12} and theoretically estimated~\cite{HozoiPreprint12,BHKim} values in \SIOsns.

\begin{table}
\begin{tabular}{c c c c c c c c c c c c c c c c c c c c}

\hline 
$J$ & & $J_c$ & & $J_2$ & & $J_3$ & & $J_{2c}$ & & $\Gamma$ & & $\Gamma_c$ & & $D$ & & $D_c$ \\
\hline
93 & & $25.2$ & & $11.9$ & & $14.6$ & & $6.16$ & & $4.4$ & & $34.3$ & & $24.5$ & & $28.1$ \\
\hline
\label{table:table1} 
\end{tabular}
\caption{Coupling constants (in units of meV) determined from fits to the experimental magnon dispersion.}
\end{table}

We now return to the discussion of the small bilayer splitting. Among the
long-range interaction terms that were included ($J_2$, $J_3$, and
$J_{2c}$), $J_{2c}$ plays a critical role in determining the bilayer
splitting. 
In a pure Heisenberg model, a small bilayer splitting would imply a small energy difference
between the two magnetic configurations shown in Fig. 3, and the
coupling constants in Tab.~1 corroborate this with $J_c\approx
4J_{2c}$ ($J_{2c}$ couples to four sites and thus cancels a four times
stronger $J_c$). However, Eqs. (4) and (5) show that the PD
interactions do not contribute to the bilayer splitting but considerably lower
the energy of the $G$-type AF order in Fig. 3(a).

The interesting situation arises that even if the inter-layer {\it isotropic} exchange interactions are almost completely
frustrated, the layers are still strongly coupled by the inter-layer PD and DM
interactions that are responsible for the magnetic anisotropy. The strong
inter-layer PD interactions manifested by the large gap do not conflict with
the small bilayer splitting since PD terms, having the Ising form, do not
propagate magnons between the planes. 

The observed large magnon gap has two important implications. First, it raises the question as to how the magnetic order melts at a temperature scale (T$_{\textrm N}$$\approx$285 K) much smaller than the magnon gap ($\Delta_m>$1000 K). 
The rapid drop in the electrical resistivity when the system is heated through T$_{\textrm N}$~\cite{cao02} suggests that 
the transport properties are correlated with the magnetic order. 
However, the observed large magnon gap can hardly be reconciled with the standard single band spin-density wave 
picture with isotropic spin dynamics. Given that the charge gap  
(even unresolved in the optical data) might be very small, it is likely that   
AF order is destroyed by thermally activated charge carriers that form
magnetic polarons, whose motion is known to be particulary detrimental for an 
Ising-type magnetic order with large magnon gap that prevents a coherent charge
propagation~\cite{ReadPRB89}. 
Whether this thermal-carrier-driven magnetic transition is a special case for
\SIO or can be generally applied to other 5$d$ TMOs with small charge
gap~\cite{Calder,Hsieh} remains to be explored both experimentally and
theoretically.  

%
%

\begin{figure}[t]
\hspace*{-0.2cm}\vspace*{-0.1cm}\centerline{\includegraphics[width=1\columnwidth,angle=0]{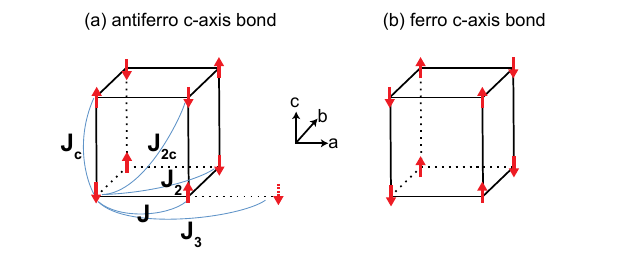}}

\caption{(a) Ground state magnetic structure of SIOns. (b) A magnetic state with nearly the same energy when only Heisenberg couplings are considered.
}\label{fig:fig3} 
\end{figure}

Second, the enhanced PD interactions suggest a direction for realization of the Kitaev model discussed in related iridates $A$$_2$IrO$_3$ ($A$=Li or Na). In an ideal geometry where Ir ions sit on the vertices of a honeycomb lattice and are connected by edge-sharing oxygen octahedra, it has been shown that the isotropic exchange interactions are strongly reduced for the cubic \jeff=1/2 wavefunction and the PD terms render a realization of the Kitaev model with 
a spin liquid ground state~\cite{jackeli09,chaloupka10}. Experimentally, however, both
Li$_2$IrO$_3$ and Na$_2$IrO$_3$ are known to have long-range order at
T$_N$$\approx$15 K~\cite{HillPRB11,SinghPRB10,SinghPRL12}, which signals
strong perturbation by the Heisenberg term. This is possibly due to
less-than-ideal realization of the Kitaev model in these compounds.  Our study
shows that approaching the MIT boundary in favor of large $J_H/U$ may enhance
the PD term and stabilize the spin-liquid ground state. In this regard, high
pressure experiments on these iridates may be interesting.  

To summarize, we have revealed the unconventional nature of the magnetism in a
spin-orbit entangled Mott insulator \SIO lying on the verge of MIT. The system
shows a marked departure from the Heisenberg model due to the strongly
enhanced PD interactions. In contrast to 3$d$ oxides with small spin-orbit
coupling that can be described by isotropic Heisenberg interactions with
small anisotropic corrections, \SIO exemplifies how a novel type of magnet can
arise from a 5$d$ oxide with strong spin-orbit coupling and a small charge
gap. Our findings should have profound implications for other iridium compounds with lattice geometries in which the Heisenberg
term is strongly suppressed.

%
%


%

\acknowledgements{Work in the Material Science Division and the use of the
  Advanced Photon Source at the Argonne National Laboratory was supported by
  the U.S. DOE under Contract No. DE-AC02-06CH11357. G.J. acknowledges support
  from GNSF/ST09-447. M.D. acknowledges support from the DFG
(Emmy-Noether program).}

\bibliography{SIO327RIXS}

\end{document}


\title{Giant Magnon Gap in Bilayer Iridate \SIO: Enhanced Pseudo-dipolar 
Interactions Near the Mott Transition}

\author{Jungho Kim} 
\author{A. Said}
\author{D. Casa}
\author{M. H. Upton}
\author{T. Gog} 
\affiliation{Advanced Photon Source, Argonne National Laboratory, Argonne, Illinois 60439, USA}
\author{M. Daghofer}
\affiliation{Institute for Theoretical Solid Sate Physics, IFW Dresden, Helmholtzstr. 20, 01069 Dresden, Germany}
\author{G. Jackeli} 
\affiliation{Max Planck Institute for Solid State Research, Heisenbergstr. 1, D-70569 Stuttgart, Germany}
\author{J. van den Brink} 
\affiliation{Institute for Theoretical Solid Sate Physics, IFW Dresden, Helmholtzstr. 20, 01069 Dresden, Germany}
\author{G. Khaliullin}
\affiliation{Max Planck Institute for Solid State Research, Heisenbergstr. 1, D-70569 Stuttgart, Germany}
\author{B. J. Kim}
\affiliation{Materials Science Division, Argonne National Laboratory, Argonne, IL 60439, USA}

\date{\today}

\begin{abstract}
\end{abstract}

\pacs{74.10.+v, 75.30.Ds, 78.70.Ck}

\maketitle

Nearest-neighbor Heisenberg couplings $J$ and $J_c$, as well as parameters 
$\Gamma$, $\Gamma_c$, $D$, $D_c$ in Eqs.~(1) and (2) of the main text, can be 
derived from a microscopic model Hamiltonian of Ref.~\cite{Jackeli:2009hz} for 
$J_{eff}=\tfrac{1}{2}$ moments (referred to as a `spin' $S$ for simplicity). 
They depend on Hund's coupling $J_H$ via
parameter $\eta=J_H/U$, on the IrO${_6}$-octahedra rotation angle $\alpha$,
as well as on the tetragonal splitting $\Delta$ parametrized 
by angle $\theta$ via $\tan 2\theta = 2\sqrt{2}\lambda/(\lambda-2\Delta)$. 
For completeness, we reproduce below the explicit expressions of  
coupling constants for $ab$-plane~\cite{Jackeli:2009hz} and 
$c$-axis~\cite{RXDpaper} bonds. They read as follows (in units of $4t^2/U$): 
\begin{align}\tag{S1}
J &= \nu_1(t_s^2-t_a^2) + \nu_2\cos^2\theta\cos 2\theta\\
J_c &= \nu_1\cos^4 \theta \cos 4\alpha\;,\tag{S2}
\end{align}
where abbreviations 
$t_s = \sin^2\theta+ \tfrac{1}{2} \cos^2\theta \cos 2\alpha$ and 
$t_a = \tfrac{1}{2}\cos^2\theta \sin 2\alpha$ are used, 
$\nu_1 = (3r_1 + r_2 + 2r_3)/6$, $\nu_2 =(r_1-r_2)/4$, with 
$r_1 = 1/(1-3\eta)$, $r_2 = 1/(1-\eta)$, and $r_3= 1/(1+2\eta)$. 
The coupling constants for the symmetric and Dzyaloshinsky-Moriya
anisotropies within the planes are 
\begin{align}\tag{S3}
\Gamma & = 2\nu_1t_a^2  - \nu_2\cos^2 \theta \cos 2\theta, \quad
D = 2\nu_1t_st_a , 
\end{align}
and
\begin{align}
\Gamma_c & = 2\nu_1\cos^4\theta \sin^2 2\alpha + 2\nu_2\cos^2
\theta\sin^2\theta, \tag{S4}\\
 D_c &= \nu_1\cos^4\theta\sin 4\alpha \tag{S5}
\end{align}
for the interlayer couplings. 

As has been pointed out for a single layer Sr$_2$IrO$_4$, in-plane canted 
spins are energetically favorable near the cubic limit $\Delta \ll \lambda$, but
larger $\Delta$ with $\theta \geq \pi/4$ induce spins to lie along the
$c$ axis~\cite{Jackeli:2009hz}. In the bilayer \SIO, the inter-layer bonds 
favor spins along the $c$-axis for any $\theta$ and thus move the 
transition to smaller $\theta$~\cite{RXDpaper}. 
We obtained magnon energies for such a magnetic pattern in linear spin-wave
theory. Due to the bilayer geometry, each in-plane momentum
${\bf q}=(q_x,q_y)$ leads to two magnon branches with
energies $\omega_{\pm}({\bf q})$ as given in Eq.~(3) of the main text.
It turns out that the large magnon gap can only be reproduced by assuming 
a rather large value for the ratio of Hund's coupling compared to onsite 
Coulomb repulsion, $\eta = J_H/U \gtrsim 0.22$. Such a larger $\eta$ 
can be motivated by noting that very small charge gap in \SIO implies more
itinerant electrons, which can more effectively screen, and thus reduce, $U$. 
Elongation along the $c$ axis also increases the magnon gap; however, 
large values of $\Delta$ would imply substantial intensities at the 
$L_2$ edge, which rises sharply for $\theta \gtrsim 0.27\pi$, and is not 
seen in experiment~\cite{RXDpaper}. This gives a constraint 
$\theta \lesssim 0.27\pi$. We find here a good fit with $\eta=0.24$ and 
$\theta =0.26\pi$, giving a nearest-neighbor in-plane $J=93\;\textrm{meV}$ 
and out-of-plane $J_c = 25.2\;\textrm{meV}$. Antiferromagnetic 
second- and third-neighbor intra-layer couplings were subsequently fitted 
to approximately reproduce the observed dispersion. In addition, an 
inter-layer second-neighbor coupling $J_{2c}$ was needed to reduce the bilayer
magnon splitting. Figure 1 shows the theoretical spectrum that closely reproduces the experimental spectrum.

\begin{figure}
\includegraphics[width=0.6\columnwidth]{Skw_kxky_v2}
\caption{Theoretical spectrum, energies according to Eq.~(3) of the
  main text 
  and intensities from Eq.~(\ref{eq:int}), depending on in-plane
  momentum $(q_x,q_y)$ for $q_z = \pi/3$.\label{fig:Skw}}
\end{figure}

Despite strong calcellation among the inter-layer Heisenberg couplings  
$J_c$ and $J_{2c}$, the layers are still rigidly coupled by the anisotropic 
pseudo-dipolar $\Gamma_c$ term which stabilizes magnetic structure shown 
in Fig.~3(a) of the main text. Being of the Ising-type, however, this 
coupling does not contribute to the splitting between the magnon branches.

\begin{figure}
\subfigure[]{\includegraphics[width=0.48\columnwidth]{Skzw_pipi_v2}\label{fig:Skzw_pipi}}
\subfigure[]{\includegraphics[width=0.48\columnwidth]{Skzw_pi0_v2}\label{fig:Skzw_pi0}}
\caption{Theoretical spectrum (\ref{eq:spec_kz}) depending on momentum $q_z$ along the
  $c$ axis perpendicular to the planes for (a) momenta $({\bf q},q_z)=(\pi,\pi,q_z)$ and
  (b) $({\bf q},q_z)=(\pi,0,q_z)$.\label{fig:Skz}}
\end{figure}

The spectral weights of the two branches are given by
\begin{align}\tag{S6} \label{eq:int}
I_{\pm}({\bf q}) =\frac{3}{8}
\frac{A_{\pm}-X_{\pm}}{\sqrt{A^2_{\pm}-X^2_{\pm}-Y^2_{\pm}}}
\end{align}
with $A_{\pm}$, $X_{\pm}$, and $Y_{\pm}$ as in Eqs.~(4) and
(5) of the main text. Since the two branches are due to the bilayer
geometry and $\omega_+$ ($\omega_-$) corresponds to a momentum
$q_z=0$ ($q_z=\pi$) along the $c$ axis, the total $q_z$-dependent
intensity is
\begin{align}\tag{S7}\label{eq:spec_kz}
I({\bf q},q_z,\omega) &\propto \cos^2 \frac{q_z}{2} I_+({\bf q})
\;\delta[\omega-\omega_+({\bf q})] \\
&+ \sin^2 \frac{q_z}{2} I_-({\bf q})\; \delta[\omega-\omega_-({\bf
q})]\;.\nonumber
\end{align}
For ${\bf q}=(\pi,\pi)$, this leads to a weak $q_z$ dependence as
weight shifts from $\omega_+$ to $\omega_-$, see
Fig.~\ref{fig:Skzw_pipi}. As there is no bilayer splitting for
${\bf q}=(\pi,0)$, no such shift occurs here, see Fig.~\ref{fig:Skzw_pi0}.